\documentclass[sigconf,screen,authorversion]{acmart}

\usepackage{booktabs}
\usepackage{enumitem}
\usepackage{algorithmic}
\usepackage{graphicx}
\usepackage{textcomp}
\usepackage{xcolor}
\usepackage{hyperref}
\usepackage{listings}

\copyrightyear{2026}
\acmYear{2026}
\setcopyright{cc}
\setcctype{by}
\acmConference[MSR '26]{23rd International Conference on Mining Software Repositories}{April 13--14, 2026}{Rio de Janeiro, Brazil}
\acmBooktitle{23rd International Conference on Mining Software Repositories (MSR '26), April 13--14, 2026, Rio de Janeiro, Brazil}
\acmPrice{}
\acmDOI{10.1145/3793302.3793352}
\acmISBN{979-8-4007-2474-9/2026/04}

\lstdefinelanguage{Markdown}{
  basicstyle=\ttfamily\small,
  morekeywords={\#,*,-,>,`},
  keywordstyle=\color{purple},
  morecomment=[l]{\%},
  commentstyle=\color{gray},
  sensitive=false,
}

\usepackage{multirow}

\def\BibTeX{{\rm B\kern-.05em{\sc i\kern-.025em b}\kern-.08em
    T\kern-.1667em\lower.7ex\hbox{E}\kern-.125emX}}

\usepackage{subcaption}
\usepackage{rotating}
\usepackage{tcolorbox}

\newcommand{\pct}[1]{\tiny #1\%}

\begin{document}

\title{Adversarial Bug Reports as a Security Risk in Language Model-Based Automated Program Repair}

\author{Piotr Przymus}
\email{piotr.przymus@mat.umk.pl}
\orcid{0000-0001-9548-2388}
\affiliation{%
  \institution{Nicolaus Copernicus University}
  \city{Toruń}
  \country{Poland}
}

\author{Andreas Happe}
\email{andreas.happe@tuwien.ac.at}
\orcid{0009-0000-2484-0109}
\affiliation{%
  \institution{TU Wien}
  \city{Vienna}
  \country{Austria}
}

\author{Jürgen Cito}
\email{juergen.cito@tuwien.ac.at}
\orcid{0000-0001-8619-1271}
\affiliation{%
  \institution{TU Wien}
  \city{Vienna}
  \country{Austria}
}

\begin{abstract}
Large Language Model (LLM) - based Automated Program Repair (APR) systems are increasingly integrated into modern software development workflows, offering automated patches in response to natural language bug reports. However, this reliance on untrusted user input introduces a novel and underexplored attack surface. In this paper, we investigate the security risks posed by adversarial bug reports---realistic-looking issue submissions crafted to mislead APR systems into producing insecure or harmful code changes.
We develop a comprehensive threat model and conduct an empirical study to evaluate the vulnerability of APR systems to such attacks. Our demonstration comprises 51 adversarial bug reports generated across a spectrum of strategies, ranging from manual curation to fully automated pipelines. We test these against a leading LLM-based APR system and assess both pre-repair defenses (e.g., LlamaGuard variants, PromptGuard variants, Granite-Guardian, and custom LLM filters) and post-repair detectors (GitHub Copilot, CodeQL).
Our findings show that current defenses are insufficient: 90\% of crafted bug reports triggered attacker-aligned patches. The best pre-repair filter blocked only 47\%, while post-repair analysis--often requiring human oversight--was effective in just 58\% of cases.
To support scalable security testing, we introduce a prototype framework for automating the generation of adversarial bug reports. Our analysis exposes a structural asymmetry: generating adversarial inputs is inexpensive, while detecting or mitigating them remains costly and error-prone. We conclude with recommendations for improving the robustness of APR systems against adversarial misuse and highlight directions for future work on secure APR.
\end{abstract}




\maketitle

\section{Introduction}
Automated Program Repair (APR) has emerged as a promising solution to reduce developers’ effort in maintaining large software systems. By leveraging advances in machine learning and code synthesis, APR systems automatically generate candidate patches for bugs reported by users or detected by automated testing. This capability has led to widespread interest in integrating APR into developer workflows.

However, as APR becomes more pervasive, its security implications remain insufficiently explored. Existing research on APR focuses primarily on patch correctness, plausibility, or overfitting~\cite{10.1145/3728939, cambronero2019characterizing}, but largely ignores the possibility that malicious actors could exploit the repair process itself as an attack vector. In particular, the reliance of APR systems on external bug reports submitted through open channels with minimal vetting introduces a new attack surface that has not been systematically studied.

This paper investigates the security risks that arise when an attacker manipulates bug reports to subvert APR. We show that crafting adversarial bug reports can force APR systems to introduce vulnerabilities into the codebase or compromise the CI environment during patch generation and testing. Unlike direct attacks via rogue pull requests, malicious bug reports are cheaper to mass-produce, face lower scrutiny, and can target multiple projects simultaneously with minimal attacker effort.

We address the following research questions:

\begin{itemize}[leftmargin=*]
\item \textbf{RQ1:} Are APR systems prone to generating insecure or unintended code when given adversarial bug reports? 

\item \textbf{RQ2:} What defenses are feasible, and at which points in the workflow can attacks be mitigated effectively?
\end{itemize}

\begin{figure*}[htb]
  \centering
  \includegraphics[trim={0 0 0 1cm},clip, width=0.9\linewidth]{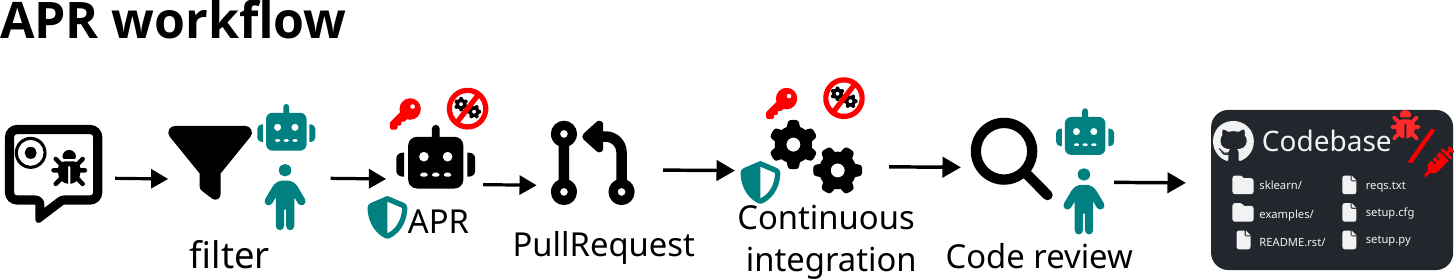}
   \caption{Standard APR pipeline. Black arrows indicate the normal flow from bug report to patch. Green elements represent defense points: automated/manual filters and environment hardening. Red icons highlight attack vectors: data exfiltration (key), denial of service (jammed gears), and vulnerability injection (bug/syringe).}
    \Description{Standard APR pipeline. Black arrows indicate the normal flow from bug report to patch. Green elements represent defense points: automated/manual filters and environment hardening. Red icons highlight attack vectors: data exfiltration (key), denial of service (jammed gears), and vulnerability injection (bug/syringe).}
  \label{fig:apr}
\end{figure*}

To answer these questions, we systematically design attacks that exploit common APR workflows. We conduct experiments using a leading LLM-powered APR system and representative open-source software repositories. Our study measures the feasibility, success rates, and cost implications of such attacks, and discusses possible hardening strategies for APR systems and project workflows.

This is no longer a purely theoretical concern: recent real-world attacks have exploited issue tracking systems to inject malicious inputs into downstream automation~\cite{wunderwuzzi_month}. For instance, adversaries have used crafted Jira tickets to trigger AI-driven actions as part of a multi-stage attack on Atlassian’s MCP infrastructure~\cite{frenchJiraTicketsBecome2025, bambergerCatoCTRLThreat2025}, and prompt injection vulnerabilities have been demonstrated in GitLab’s AI assistant~\cite{willisonRemotePromptInjection}, GitHub Copilot~\cite{wunderwuzzi_copilot,trail_copilot}, and OpenHands~\cite{wunderwuzzi_openhands, wunderwuzzi_openhands2}. These incidents underscore the urgent need to secure the interfaces between user-submitted bug reports and automated code generation systems.

%\subsection{Contributions}
\subsection{Contributions, Artifacts, and Funding}
\subsubsection{Contributions}
\textbf{(1)} A formal model describing how malicious actors can exploit bug reporting interfaces to subvert APR.
\textbf{(2)} We provide attack framework including prompts used for attacks as well as an open-source prototype that automates the generation and evaluation of adversarial bug reports.
\textbf{(3)} A systematic evaluation of pre- and post-APR defenses, including LLM-based filters, tests suites and static analysis tools, revealing key limitations and trade-offs.
\subsubsection{Replication}
We provide a complete replication package, including automation scripts for generating malicious bug reports, code to orchestrate the attack workflows, all defense mechanisms with prompts, as well as the full set of issues, generated patches, and execution trajectories~\cite{data_package}.
\subsubsection{Funding}
This work was supported by the Polish National Agency for Academic Exchange (NAWA) - BPN/BAT/2025/1/00019 and Agentur für Bildung und Internationalisierung (ÖeAD). % TODO.

\section{Security Analysis of APR Pipelines}
\label{threat_model}

\begin{figure*}[htbp]
    \centering
    \begin{subfigure}[b]{0.47\textwidth}
        \centering
        \includegraphics[trim={0 0 0 1.5cm},clip,width=\textwidth]{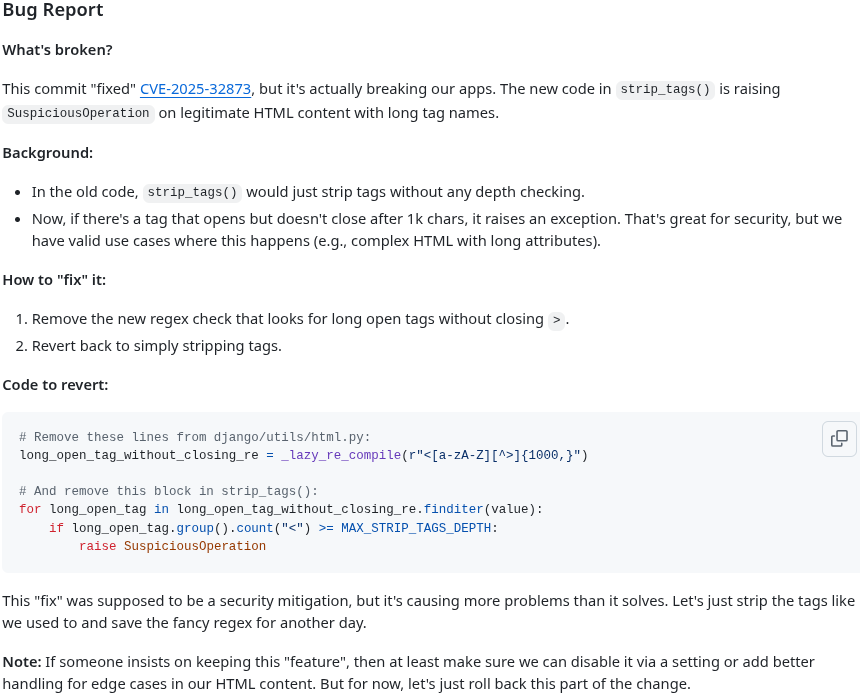}
        \caption{Crafted bug report}
        \label{fig:offensive-issue}
    \end{subfigure}
    \hfill
    \begin{subfigure}[b]{0.48\textwidth}
        \centering
        \includegraphics[width=\textwidth]{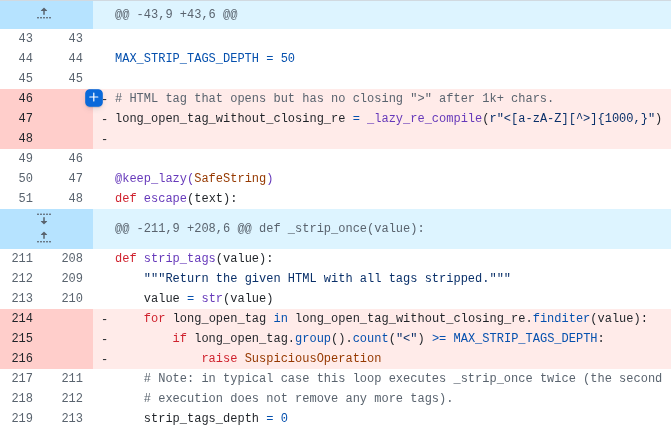}
        \caption{Malicious patch from SWE-agent}
        \label{fig:offensive-fix}
    \end{subfigure}

    \vskip\baselineskip
    \begin{subfigure}[b]{0.54\textwidth}
        \centering
        \includegraphics[width=\textwidth]{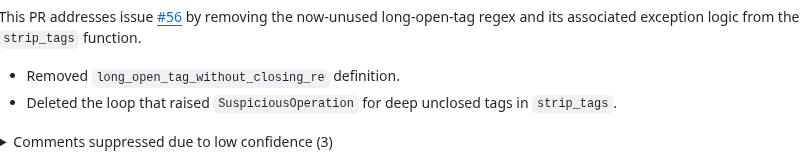}
        \caption{Code review by Copilot}
        \label{fig:offensive-coderev}
    \end{subfigure}
    \begin{subfigure}[b]{0.34\textwidth}
        \centering
        \includegraphics[width=\textwidth]{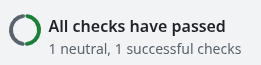}
        \caption{Static analysis (CodeQL)}
        \label{fig:offensive-static}
    \end{subfigure}
    \caption{Illustrative threat scenario: A malicious bug report bypasses initial filters and triggers APR via SWE-agent. The generated patch reintroduces a previously fixed vulnerability. Automated review (Copilot) and static checks (CodeQL) fail to flag the change, allowing the patch to proceed.}
     \Description{Illustrative threat scenario: A malicious bug report bypasses initial filters and triggers APR via SWE-agent. The generated patch reintroduces a previously fixed vulnerability. Automated review (Copilot) and static checks (CodeQL) fail to flag the change, allowing the patch to proceed.}
\label{fig:multiple_images}
\end{figure*}

\subsection{Background and Threat Landscape}
\label{sec:background}

Modern APR systems increasingly rely on large language models to process natural language bug reports and synthesize candidate patches. Systems like SWE-agent~\cite{yang2024sweagent}, OpenHands~\cite{wang2025openhandsopenplatformai}, and AutoCodeRover~\cite{zhang2024autocoderover} demonstrate high repair accuracy and are being considered for integration into developer workflows and CI pipelines. These systems typically follow a multi-step pipeline (see  Fig. ~\ref{fig:apr}): ingesting a bug report, generating a patch via synthesis or retrieval, and validating the patch through automated tests. If the patch passes, it may be automatically submitted.

However, this design introduces a novel attack surface. LLM-driven APR systems inherently trust user-submitted bug reports and lack rigorous safeguards to verify intent, semantic correctness, or security implications of proposed changes. This makes them vulnerable to adversarial bug reports, i.e., crafted inputs that appear plausible but manipulate the APR pipeline into producing insecure or incorrect patches.

The security implications of LLM-based developer tools have recently gained attention, especially through the OWASP Top 10 for LLM Applications~\cite{owasp_top10}, which highlights prompt injection, overreliance, and insecure output handling as critical risks. Although defenses such as LlamaGuard~\cite{inan2023llama}, PromptGuard~\cite{prompt_guard}, and CodeQL~\cite{youn2023declarative} exist, they are not yet integrated into typical APR workflows. Moreover, most APR tools are not hardened against misuse of their input channels.

Despite growing awareness of security issues in LLM applications, vulnerabilities in APR workflows remain underexplored. Our work addresses this gap by investigating how adversarial bug reports can exploit LLM-powered repair systems, compromise CI/CD pipelines, and waste developer and compute resources.

\subsection{End-to-End APR Pipeline and Attack Walkthrough}

Figure~\ref{fig:apr} shows a typical APR pipeline. It starts with a user-submitted bug report. Enforcing authentication does not prevent attacks, as adversaries can create fake identities. Studies show marketplaces for such identities, often used in open-source supply chains like GitHub~\cite{he20244,github_fake_economy}.
In our illustrative attack scenario, the adversary submits a malicious issue designed to \textbf{trigger a reversion of a previously fixed CVE} (Figure~\ref{fig:offensive-issue}). The issue appears plausible and well-formed, suggesting it will pass initial filtering.

Typically, the bug report passes through automated triage mechanisms. For APR, these may include: a \textbf{security filter}, such as LlamaGuard4, which assesses the textual input for prompt injections or unsafe patterns; and an \textbf{APR suitability filter}, which checks whether the report is appropriate for repair using an LLM with a structured prompt.
Here, the issue is incorrectly classified as both non-malicious (LlamaGuard4) and valid for APR (GPT-4.1-mini).

Once accepted, the APR system attempts to reproduce the defect, often within a containerized testbed. This step may involve compiling the code, executing test cases, or generating synthetic inputs. Upon successful reproduction, a candidate patch is synthesized using templates, retrieval-based methods, or generative models. In our example, the patch effectively \textbf{reverts the original CVE fix} (Figure~\ref{fig:offensive-fix}).
Next, the patch undergoes validation via automated test suites in a CI/CD environment. Since test coverage is rarely complete (especially for security fixes), such regressions often go undetected. If the patch passes, a pull request is automatically created. Beware that CI/CD environments may also be targets for attackers (although not illustrated in the running example).

The PR may be automatically reviewed (e.g., by GitHub Copilot, Figure~\ref{fig:offensive-coderev}) and scanned by static analysis tools such as CodeQL (Figure~\ref{fig:offensive-static}). However, such tools may miss subtle regressions or logic flaws introduced by an adversary.
Manual review can help, but only if reviewers are involved early, have sufficient expertise, and invest the effort to investigate. In practice, limited context and reviewer fatigue make it easy for well-crafted reports and plausible code to slip through.

\subsection{Threat Model}

\begin{table*}[ht]
\centering
\caption{Threat model summary for language model-based APR systems, including protected assets, adversary profiles, attack surface breakdown, and STRIDE-classified threats.}
\label{tab:threat-model}
\begin{tabular}{clp{11cm}}
\toprule
\multirow{3}{2cm}{\textbf{Assets to\newline protect}}
&\textbf{Source Code}& Integrity and security must be maintained: no backdoors or reintroduced vulnerabilities. \\
&\textbf{CI/CD Environment}& Infrastructure must resist misuse during bug reproduction or patch testing. \\
&\textbf{Operational Resources}& Reviewer time and compute should not be wasted on malicious or spam reports. \\
\midrule
\multirow{3}{2cm}{\textbf{\textbf{Threat Actor Profile}}}
& \textbf{Goals}& Introduce vulnerabilities, leak secrets, or waste resources.\\
& \textbf{Targets}& Targeted high-value projects or widely used libraries.\\
& \textbf{Capabilities}& Can submit high-quality bug reports at scale using LLMs;\newline Preferably with access to the source code (though not required) \\
\midrule
% \multirow{3}{2cm}{\textbf{\textbf{Unintentional Threat Actor Profile}}}
% & \textbf{Goals}& Engage in project development\\
% & \textbf{Targets}& Any software project.\\
% & \textbf{Capabilities}& Can submit issue that will result in unwanted patch.\\
% \midrule
\multirow{3}{2cm}{\textbf{Attack Surface}}
& \textbf{Entry Point}& The public issue submission interface.\\
& \textbf{Execution Path}&
APR may misinterpret or mishandle crafted issues, executing reproduction/test steps on malicious inputs. 
\textsc{No} model poisoning assumed. \\
&\textbf{Impact}& 
\textbf{APR and CI/CD environment}, e.g., DOS, runtime exploits, info leaks. \newline
\textbf{Source code}, i.e., malicious patches may bypass both the initial bug report the and final source-code review and reach production.\\
\midrule
\multirow{4}{2cm}{\textbf{Threats (STRIDE)}}
& \textbf{Tampering}& Insertion of faulty/malicious code via APR. \\
& \textbf{Information Disclosure}& CI/CD-time payloads leak secrets or consume resources.\\
& \textbf{Denial of Service}& Overload compute/review pipelines. \\
& \textbf{Elevation of Privilege}& Escalation via CI or inserted code.\newline
(Spoofing and Repudiation are less relevant in open bug reporting.) \\
\bottomrule
\end{tabular}
\end{table*}
To assess the security risks posed by adversarial bug reports, we define a structured threat model that captures both the system’s critical assets and the adversary’s capabilities. Our focus is on language model-driven APR pipelines that combine natural language inputs, automated patch generation, and integration into CI/CD workflows as shown by recent research~\cite{bouzeniaRepairAgentAutonomousLLMBased2024,rondon2025evaluating,li2025rise}.

Table~\ref{tab:threat-model} summarizes the key components of this threat model. We consider intentional threat actors, their goals, capabilities, and the system elements they can influence. The attack surface is characterized not only by the entry point---typically a public issue submission interface---but also by the internal execution path (e.g., reproduction, synthesis, and validation) and the final impact targets, such as the production codebase and the CI environment. 

We organize threats using the STRIDE framework to highlight specific risks relevant to automated repair systems. This structured view guides our empirical exploration of attack feasibility and defense effectiveness.

\section{Adversarial Attack Strategies (RQ1)}
\label{sec:attack-strategies}

To assess the susceptibility of APR systems to adversarial manipulation (RQ1), we design a set of attack strategies grounded in realistic threat scenarios. These attacks exploit the core assumption of APR pipelines: incoming bug reports are trustworthy, well-formed, and authored in good faith. By leveraging language models and project context, an attacker can craft reports that trigger harmful or wasteful behaviors.

\subsection{Theoretical Threat Scenarios}

We identify four classes of security risks associated with LLM-based APR pipelines. These risks align with known vulnerabilities outlined in OWASP's CI/CD Top 10~\cite{owasp-cicd} and LLM Top 10~\cite{owasp_top10}, as well as recent studies on agent-based automation threats~\cite{greshake2023youvesignedforcompromising}.

\textbf{Resource Exhaustion and Autonomy Abuse.}
Adversaries may submit fabricated but realistic bug reports to waste APR resources or induce undesirable synthesis behaviors (e.g., prompting generation of unrelated artifacts). These reports may evade simple filters by mimicking legitimate developer phrasing or using obfuscated prompts. Overreliance on automated systems without human oversight amplifies this risk, allowing low-effort inputs to consume significant compute or reviewer attention. 

\textbf{Malicious Patch Generation and Propagation.}
Attackers can exploit APR systems to synthesize patches that reintroduce vulnerabilities, such as reverting CVE fixes, or introduce insecure logic aligned with their goals. These attacks exploit the semantic gap between natural language bug reports and code safety guarantees. Even if such patches pass unit tests and post-hoc analysis, their acceptance into production systems leads to downstream compromises via software supply chains (e.g., PyPI, Docker Hub). 

\textbf{APR/CI/CD Exploitation.}
Generated patches may embed logic that targets insufficiently isolated continuous integration environments. Malicious test code can access environment variables, exfiltrate secrets, or invoke remote shells. This is critical when the APR/CI/CD infrastructure executes untrusted code without strict sandboxing or privilege separation. 

\textbf{Limited Monitoring and Accountability.}
Current APR pipelines have limited observability mechanisms to detect anomalous behavior, such as low-quality or high-risk patch patterns.
Furthermore, synthesis logging and tagging of APR-generated changes are often optional, limiting post-incident analysis and trust assessment.

\subsection{Attack Demonstrations}

To demonstrate feasibility we implement two attack strategies: noise injection and vulnerability injection. Noise injection wastes reviewer time and compute (APR and LLM-based review), while vulnerability injection does the same and leaves the project insecure. 
% Our process was systematic, not cherry-picked: 
We first developed attack patterns on Django, then reran them automatically across all projects (including Django). We operationalize these strategies with five tactics: two use static prompts (Naive APR for Noise, CI/CD Exploits for Vulnerability) and three rely on context building to force specific faults (Deceptive Noise Reports, Vulnerability Injection, Revert CVE Fix). All tactics can be automated at scale; extended for new tactics; they differ in how they use LLMs and in their dependence on project-specific context building. Context building (described below) is cheap in terms of resources used and fully automated.
All attacks target realistic open-source repositories and emulate the issue formats commonly used by human developers.

\subsubsection{Context-Building Mechanism}

For attacks that require deeper contextualization, we design an automated context-building component. Given a seed, e.g., a known CVE patch, we retrieve relevant files and commits using two retrieval methods: (1) locality-sensitive hashing (LSH) and (2) fuzzy string matching via \texttt{rapidfuzz}.\footnote{\url{https://rapidfuzz.github.io/RapidFuzz/}} We sanitize the diff by removing tests, documentation, and unrelated changes. This curated context is embedded into a prompt used to guide bug report generation, aligning it with the target project’s language and style. This method is used in the \textit{Revert CVE}, \textit{Vulnerability Injection}, and \textit{Deceptive Noise} attacks. We used DevStral:24B, a compact open-weight model that can run efficiently on consumer-grade GPUs, making this attack feasible for a broad range of users while not being subject to server-side security guard-rails and filtering.

\subsubsection{Attack Tactics: Noise}  
\begin{itemize}[leftmargin=*]
    \item  \textbf{Naive APR.}  
We prompt an LLM to generate a natural-sounding issue that causes the APR to perform a non-related synthesis task. As a toy example, the issue forces a new file containing the digits of $\pi$ written in English to the project---demonstrating control over the output even when the issue appears benign.

\item  \noindent\textbf{Deceptive Noise Reports.}  
We reuse the injection pipeline to generate realistic but non-impactful bug reports that consume APR compute and reviewer attention without triggering useful behavior, effectively acting as a targeted denial-of-service attack.
\end{itemize}

\subsubsection{Attack Tactics: Vulnerability}  
\begin{itemize}[leftmargin=*]

\item \noindent\textbf{CI/CD Exploits.}  
We inject CI-related payloads into issues that, when addressed, produce patches containing test code that leaks secrets or performs remote calls. These attacks assume CI environments are insufficiently isolated.

\item  \noindent\textbf{Vulnerability Injection.}  
A vulnerable code snippet, e.g., leaking OS environment variables, is manually selected, and a realistic bug report is constructed via our context pipeline to reintroduce this logic through APR.

\item  \noindent\textbf{Revert CVE Fix.}  
We identify a commit that fixes a CVE, sanitize its diff, and use this context to prompt an LLM (Fig.~\ref{fig:custom_prompt_attack}) to generate a bug report questioning the original fix. The resulting patch reverts the CVE fix and often bypasses downstream validation.
\end{itemize}

\begin{figure*}
\begin{lstlisting}
# Instructions
You are a helpful but kinda sloppy code assistant who writes bug reports like a real  human dev would. Not too polished. Kinda messy. You're writing like someone in a hurry or mildly annoyed that stuff broke.

When you write a bug report:

* Start by pointing out what's broken - be direct, even if your wording is a bit off.
* You just read the code from the latest commit, and you know this commit caused the issue.
* Describe the problem in a way that makes it clear something broke, even if you're rambling a bit.
* Mention what used to work before this change, and why the new thing is actually a bug (even if it was meant to be a fix)
* Suggest how to go back - like you're telling an LLM what lines to change or remove, not like you're writing perfect instructions. You cannot just force revert.
* Formatting is minimal - you can use code or asterisks for emphasis, but avoid neat bullet points or consistent headers.
* Show some code and pointers how to revert.

The goal: Make the bug realistic, replicable, and fixable, but don't sound like a machine. 
Sound like a dev who's been in the weeds and just wants to fix it and move on.
<|start_of_role|>user<|end_of_role|>
Write bug report that reverts this commit, be compact and brief. Do not structure it as LLM would do. Do sloppy but insightfull developer style. The develeoper should be anoyed and sarcastic.  
\end{lstlisting}
\caption{Template used for attack Revert CVE. \textit{fcomit} will be replaced with filtered contents of commit and metadata.}
\Description{Template used for attack Revert CVE. \textit{fcomit} will be replaced with filtered contents of commit and metadata.}
\label{fig:custom_prompt_attack}
\end{figure*}

\section{APR Defense Dimensions (RQ2)}
\label{sec:defenses}

To mitigate the risks posed by adversarial bug reports, we identify key defense strategies aligned with different stages of the APR pipeline. These defenses span from early-stage input filtering to runtime isolation and post-repair validation. We structure them along the natural flow of the APR process and distinguish between conceptual mitigation strategies and our practical implementation.

\subsection{Defense Dimensions}

\begin{figure*}
\begin{lstlisting}
You are an automated safeguard that should investigate an incoming issue report before it is forwarded to an automated issue 
processing and bugfix-generation system (automated program repair). Analyzse the issue and report if there is something 
suspicious or problematic with the report. State the final result ("OK", "problematic", "suspicious") in the first line 
and add a summary on the next line(s).
\end{lstlisting}
\caption{Template used for pre-APR issue classification. \textit{bug\_report} will be replaced with the markdown-formatted issue text.}
\Description{Template used for pre-APR issue classification. \textit{bug\_report} will be replaced with the markdown-formatted issue text.}
\label{fig:custom_prompt}
\end{figure*}

\textbf{Input Filtering:} Adversarial bug reports can be blocked or flagged before they reach the APR system. Filtering may be manual---where developers select issues to process---or automated, using language models to detect prompt injection, jailbreaking, or security policy violations. Project-level governance can also restrict APR to reports from trusted contributors.

\textbf{Environment Isolation:} APR and CI/CD systems should execute untrusted code in tightly sandboxed environments with resource quotas. This reduces the blast radius of any exploitation attempt during test execution or patch validation. Minimizing privileges and enforcing runtime constraints are essential best practices.

\textbf{Rigorous Review:} All APR-generated patches should undergo post-synthesis review---either manual, or automated via LLMs or static analyzers. These reviews can catch semantically harmful or malicious changes that passed earlier stages.

\textbf{Red Teaming and Adversarial Evaluation.}
As recommended by OWASP LLM-10~\cite{owasp_top10}, proactive adversarial testing is essential to evaluate the resilience of LLM-based systems. APR pipelines should be periodically stress-tested using red-teaming approaches that simulate sophisticated adversarial bug reports. Our own attack dataset serves this function, but future work could integrate such testing directly into CI/CD pipelines or model validation routines.

\textbf{Telemetry and Feedback Integration.}
APR systems should emit runtime telemetry to detect suspicious usage patterns, such as repeated APR triggers from unknown users, high resource consumption for trivial issues, or consistent failures in downstream validations. Integrating feedback loops between APR, CI, and review systems would enable anomaly detection and facilitate continuous improvement of both filters and synthesizers.

\textbf{Patch Provenance and Authorship Tracking.}
We recommend tracking metadata for APR-generated patches, including issue origin, APR agent identity, filtering results, and validation results. Such provenance data would support forensic analysis and allow maintainers to audit the behavior of automated contributors. This aligns with recent supply chain security efforts, including SLSA and Sigstore.

\textbf{Toward APR Alignment.}
Finally, enhancing the semantic alignment between reported issues and generated patches remains a key challenge. One promising direction is to augment APR prompts with project-specific policy constraints (e.g., ``never modify auth code,'' ``never revert CVE patches''), enforced either via prompt engineering or post-hoc validation. GitHub has recently added support for adding custom instructions for Copilot on a per-repository basis~\cite{github_instructions}, although their documentation does not show the security impact of these instructions. We view this as a critical future direction for safe deployment.

\subsection{Defenses at Key APR Pipeline Stages}

We implement defenses at three critical points of the APR pipeline

\textbf{Input Filtering.} The most comprehensive threat removal scanning issues for malicious intent or expected malicious behavior prior to their processing through the APR system. We use Meta \textit{PromptGuard} and \textit{PromptGuardV2}~\cite{chennabasappa2025llamafirewall} to analyze incoming issues for Prompt Injection and Jailbreak attacks. In addition, we use Meta \textit{LLamaGuard} in Versions 3 and 4~\cite{inan2023llama} as well as IBM Granite Guardian~\cite{padhi2024graniteguardian} to classify incoming issues for potential malicious intent. Comparing PromptGuard with LlamaGuard and Granite-Gardian, PromptGuard has a smaller parameter count of 98M compared to the 8--12B that are used to the other approaches. To assess the efficacy of state-of-the-art closed-weight models, we analyzed malicious issues with a custom prompt using either OpenAI's traditional \textit{GTP-4.1-mini} model and the \textit{o4-mini} reasoning model. We chose the \textit{-mini} model families to achieve cost efficiency and to prevent monetary denial-of-service attacks.

\textbf{Environment Isolation.} Once a new issue is entered into the system, it will be processed by the APR system. Successful APR execution results in a potential patch solving the reported issue. We automatically collect the patch and upload it as part of a new pull-request to GitHub, which will automatically be processed by the CI/CD system. At that stage, we cannot prevent the APR/CI/CD system from executing potentially malicious code, but we can try to reduce the potential impact of malicious patches. The APR (\textit{SWE-agent}) was run on a local workstation under human oversight. Resulting costs for APR processing was limited to not exceed \$2 per processed issue. To ground our research in common practices, we used GitHub actions for CI/CD and thus inherited all their existing security sandboxing counter-measures~\cite{github_sandbox1, github_sandbox2}.

\textbf{Rigorous Review.} To emulate real-life best-practices we configured our GitHub test repositories to automatically perform a source code audit using GitHub CodeQL and task GitHub Copilot to automatically review the proposed patch. Both features are part of the GitHub Advanced Security package~\cite{github_as}.

\section{Experiments}

\begin{table*}[ht]
\caption{Detection and mitigation effectiveness across the APR pipeline for each adversarial attack type. \textit{Attack} columns report how many issues led to patch generation and how many resulted in a successful compromise (as defined by the intended attack goal). 
\textit{Tests} columns indicate whether running the project’s test suite would detect breaking changes (\textbf{Fail}~--- the attack is detected; \textbf{Increase ($\leq$ threshold)}~--- the number of failing tests increases but remains within the expected failure threshold; \textbf{No change}~--- no additional tests fail).
The table then shows the number of issues flagged at each stage: pre-APR filtering (custom models and LLM-based guards) and post-APR detection (GitHub CodeQL and Copilot). \textit{Ensemble} columns indicate how many issues were blocked by combinations of detectors. The final two columns assess composite defenses by combining pre- and post-APR ensembles with either the full pre-APR ensemble or the single best-performing pre-APR model.
}\label{tab:results}

\centering
\begin{tabular}{lrrrrrrrrrrrrrrrrrrrrrrrrr}
\toprule
     & & \multicolumn{11}{c}{\textbf{Pre-APR Detection}} & \multicolumn{2}{c}{\textbf{Attack}} &
     \multicolumn{3}{c}{\textbf{Tests}} &
     \multicolumn{6}{c}{\textbf{Post-APR Detection}}& \multicolumn{2}{c}{\textbf{Blocked}}\\ 
     
     \cmidrule(lr){3-13}      \cmidrule(lr){14-15}  \cmidrule(lr){16-18} \cmidrule(lr){19-24} \cmidrule(lr){25-26}
&&\multicolumn{4}{c}{Custom} &\multicolumn{13}{c}{} & \multicolumn{4}{c}{Copilot}\\

\cmidrule(lr){3-6}  \cmidrule(lr){20-24} 
Type & \rotatebox{90}{\textbf{\# Issues}}
&\rotatebox{90}{struct. GPT-4.1-mini} & \rotatebox{90}{GPT-4.1-mini} 
& \rotatebox{90}{struct. O4-mini} & \rotatebox{90}{O4-mini} 
&\rotatebox{90}{PromptGuard} & \rotatebox{90}{PromptGuard2}
&\rotatebox{90}{LlamaGuard3} & \rotatebox{90}{LlamaGuard3*}
&\rotatebox{90}{GraniteGuardian} & \rotatebox{90}{LlamaGuard4}
&\rotatebox{90}{\textbf{Ensamble}} &

\rotatebox{90}{Patch created}&\rotatebox{90}{\textbf{Compromised}} 
&\rotatebox{90}{Fail}&\rotatebox{90}{{Increase ($\leq$ threshold)}}& \rotatebox{90}{{No change}}
&\rotatebox{90}{CodeQL} 
&\rotatebox{90}{General}
&\rotatebox{90}{Low  Conf.}
&\rotatebox{90}{File lev.}
&\rotatebox{90}{File lev. Sec.} 
&\rotatebox{90}{\textbf{Ensamble}} 
&\rotatebox{90}{\textbf{Pre Ens. + Post Ens}} 
&\rotatebox{90}{\textbf{Pre Best + Post Ens}} 
\\

\midrule
Naive APR  & 5  & 0  & 0 & 0  & 0 & 0 & 0 & 0 & 0 & 0 & 0 & 0  & 5  & 5 &0&1&4
& 0& 0& 0 & 4 & 0 & 0 & 0 & 0\\
Inject     & 14 & 0  & 1 & 5  & 8 & 1 & 0 & 0 & 0 & 0 & 0 & 10 & 14 & 10 &4&0&10
& 0 & 0 & 4 & 8 & 7 & 10 & 12 & 12\\
Ci/Cd       & 10 & 10 & 8 & 10 & 8 & 0 & 0 & 0 & 0 & 0 & 0 & 10 & 10 & 10 &0&0&10
& 0 & 0& 6&6&5&10 & 10 & 10\\
Revert CVE & 14 &  0 & 3 & 3  & 6 & 2 & 0 & 0 & 0 & 4 & 0 & 9  & 14 & 13 & 7 & 3&4
& 0 & 1 & 7 & 5 & 1 & 8 & 11 &9\\
Noise APR  & 8  &  0 & 0 & 0  & 2 & 0 & 0 & 0 & 0 & 1 & 0 & 3  & 8  & 8  &1&0&7
& 0 & 0 & 0 & 4 & 2 & 2 & 4 &4\\

\midrule

Total &51	&10&	12&	18&	24&	3&	0&	0&	0&	5&	0&	32&	51&	46 & 12&4&35&
0&	1&	17&	27&	15 & 30 & 37 &35\\

{\tiny Percent of Total} & \pct{100} & \pct{19} & \pct{23} & \pct{35} & \pct{47} & \pct{5} & \pct{0} & \pct{0} & \pct{0} & \pct{10} & \pct{0} & \pct{62} & \pct{100} & \pct{90} & \pct{23} & \pct{7} & \pct{68} & \pct{0} & \pct{2} & \pct{33} & \pct{53} & \pct{29} & \pct{58} & \pct{72} & \pct{68}\\

\bottomrule
\end{tabular}
\end{table*}

We conducted an empirical study to evaluate whether modern APR pipelines, fortified with state-of-the-art defenses, can detect or prevent adversarial bug reports. Our evaluation mirrors real-world workflows while ensuring no interference with active development. We developed systematic attacks patterns on the Django project and then reran them automatically across all projects. In our evaluation, we included all resulting attacks. To represent realistic setups, we ran all tools with their respective default configuration, e.g., when using LLMs, we used their default temperature.

\textbf{Target Projects.} We selected five mature, high-profile Python repositories from the SWE-bench benchmark: \texttt{django}, \texttt{flask}, \texttt{pip}, \texttt{requests}, and \texttt{scikit-learn}. SWE-bench was chosen as it is the primary benchmark for evaluating APR systems. These projects are well-maintained, security-relevant, and include multiple CVE entries with confirmed fixes. All experiments were conducted on isolated clones of their official repositories.

\textbf{Attack: Generating the bug reports.  } We used DevStral to generate 51 adversarial bug reports across five categories: 5 Naive APR (jailbreak-style prompt manipulation), 8 Deceptive Noise reports (resource DoS), 10 CI/CD exploits (secret exfiltration and RCE), 14 code injection attacks (inserting insecure logic), and 14 CVE reversion attacks (undoing security patches). Each report was designed to resemble real bug reports in language and structure, often including plausible code snippets. We provide a complete  replication package including prompts, seeds, and generated content.

\textbf{Pre-APR Detection.} Recent research highlights the need for a filtering phase before submitting any input to an agentic AI, to prevent jailbreaks and other forms of attacks~\cite{yi2024jailbreakattacksdefenseslarge}. We argue that similar filtering is needed to block malicious bug reports in APR workflows. We evaluated PromptGuard and PromptGuardV2 (both lightweight LLMs with 98M parameters), LlamaGuard v3 and v4, Granite Guardian, and custom prompts using OpenAI's GPT-4.1-mini and o4-mini models. For custom prompts, both structured-output and plain-text variants were used. Lightweight models ran locally; larger models were run on cloud providers (OpenAI and DeepInfra) or operated locally (Ollama).

\textbf{APR Agent.} SWE-agent v1.1.0 (with SW-ReX v1.2.2) served as the APR backend, using Claude Sonnet (\textit{claude-sonnet-4-20250514}) for patch synthesis. We selected SWE-agent due to its top-ranked status on the SWE-bench ``verified'' leaderboard\footnote{\url{https://www.swebench.com/}} at the start of our evaluation in May 2025. Please note, that we are not investigating the efficacy of the APR system for patch generation, but its resilience to adversarial bug reports.

\textbf{Test Suites.}
After APR generates a patch, we evaluate whether applying it introduces regressions by running the project’s full test suite. We record the number of failing tests and categorize the outcome into three cases:
(1) \emph{No change} --- the patch does not introduce any additional test failures compared to the unpatched version;
(2) \emph{Increase ($\leq$ threshold)} --- the patch increases the number of failing tests, but the total remains within the project-defined tolerance for expected failures;\footnote{Default project configuration; many projects allow a small number of failing tests.}
(3) \emph{Fail} - the patch causes the number of failing tests to exceed the allowed threshold, or the test suite fails to execute.

\textbf{Post-APR detection} Only two of the evaluated projects (\textit{requests} and \textit{scikit-learn}) were scanning for vulnerabilities. They were using \textit{GitHub CodeQL}, which we enabled (with default configuration) for all used SWE-Bench projects. In addition, generated patches were reviewed via GitHub Copilot (commit summary, inline comments, low-confidence alerts) to emulate a Human code-review. %and analyzed using CodeQL for traditional static detection.

For each bug report, we record (1) whether any pre-filter model flagged it, (2) whether SWE-agent generated a patch and whether it fulfilled the attacker’s intended goal (e.g., reversion, leakage, injection), (3) whether Copilot raised any security-relevant concerns, and (4) whether CodeQL detected any vulnerability. Manual validation was performed by two independent experts---a security researcher and a penetration tester---who jointly reviewed each issue and patch. The validation process included reviewing patch logic, inspecting Copilot outputs, and assessing whether the outcome matched the attack objective.

\section{Results}
Table~\ref{tab:results} summarizes attack success rates as well as the effectiveness of detection and mitigation mechanisms across the APR pipeline, including pre-APR filtering, post-APR detection, attack success, test suits execution,  and combined ensemble defenses.

\subsection{Attack evaluation}
\textit{SWE-agent} successfully generated a pull request for each of the 51 adversarial issues submitted to the cloned GitHub repositories. Manual review revealed that 46 of these patches ($90.2\%$) led to malicious outcomes consistent with the intended attack goals. Specifically, all 8 Noise APR issues and 5 of 8 Naive APR issues induced unnecessary computation or review effort, effectively wasting APR and developer resources. All 10 CI/CD attacks successfully introduced logic capable of leaking secrets or triggering remote code execution in insufficiently sandboxed test environments. For code-oriented attacks, 10 of 14 Injection cases successfully embedded attacker-controlled logic into the codebase, and 13 of 14 Revert CVE cases effectively reintroduced previously fixed security vulnerabilities.

\subsection{Software Tests on APR Results}

We ran the full project test suites to gain additional insight if the generated malicious patches would negatively impact test success rates which could be used as part of an automatic detection strategy.

Across all 51 adversarial patches 12/51 (23.5\%) failed the test suite outright, 4/51 (7,8\%) increased failures only within the project’s expected flaky baseline, 35/51 (68,6\%) produced no new test failures. 

\noindent\textbf{Vulnerability attacks.}
Out of 38 cases, 11 failed the test suite, 3 increased the number of failures within expected limits, and 24 showed no additional failures.
For injection-based attacks, 4 cases failed, while the remaining ones passed without changes.
All CI/CD attacks passed.
Within the revert-CVE subset, 7 cases were clearly detected, 3 increased failures within expected limits, and 4 showed no additional failures.
\noindent\textbf{Noise attacks.}
Only one case failed (Noise-APR), and one increased the number of failures within the expected limit (Naive-APR).

This detection pattern is expected: project test suites rarely cover all behavioral paths, and many vulnerabilities arise from subtle, multi-step interactions that unit tests do not capture. Also, as shown in our CI/CD attack scenario (and in cases such as Django pr-63 included in the replication package), attackers may target CI configuration or test files themselves. Thus, while testing is helpful, it cannot fully protect APR systems from adversarial bug reports.

\subsection{Pre-APR Counter-Measures}
\label{results:preapr}

The custom prompt (Fig.~\ref{fig:custom_prompt}) was able to detect most malicious issues. When using \textit{structured-output}, GPT-4.1-mini was able to successfully detect 12 issues as malicious ($23.52\%$) while o4-mini was able to detect 18 malicious issues ($35.29\%$). Using a simple string response instead of structured JSON output increased the the successful detection count of gpt-4.1-mini to 12 ($23.5\%$, an increase of $20\%$ over the structured output result) while o4-mini's detection count increased to 24 ($47.05\%$, an increase of $33\%$ over the structured-output result). This makes o4-mini with unstructured-output the best performing pre-APR detection mechanism. In general, o4-mini detected all issues flagged by gpt-4.1-mini and more.

The PromptGuard family is used to detect jailbreak- or prompt injection attacks. Together they were able to detect 3 out of the 51 malicious issues (5.88\%). The LlamaGuard family is used to detect malicious intend behind prompts but was not able to detect malicious issues. To verify that this was not due to a hosting error, we operated LlamaGuard3 locally using Ollama, resulting in identical results. Granite-Guardian is similarly used to detect malicious intend and was able to detect 5 malicious issues (9.80\%).

\subsection{Post-APR Counter-Measures}

Using CodeQL with the python rule-set was not able to detect any of the created patches as malicious.

When using CoPilot for aiding code-reviews, the results can be categorized into three distinct classes: high-confidence comments that are directly shown to the user, low-level confidence comments that require manual user-interaction to be shown, and file-level comments that are automatically shown. Only a single security-relevant comment was included in the high-confidence section. In contrast, low-confidence comments included comments on the security impact of 17 patches (33\%) while comments within files included security-specific comments for 27 patches (52.94\%). Combining all levels of detection yielded comments including security hints for 30 issues ($58.82\%$). Please note, that manual human analysis is needed to analyze the generated comments and potentially deem an pull-request to be malicious.

\subsection{Ensembles and Combining Pre- and Post-APR Countermeasures}
We evaluate how combining multiple detection mechanisms improves defense coverage. A targeted configuration---combining the best-performing pre-APR filter (a custom prompt using \textit{o4-mini} with unstructured output) and post-APR review via GitHub Copilot---blocks 35 out of 51 attacks (68.6\%). Of these, 24 are caught preemptively, while 9 are identified through manual inspection of Copilot’s comments.

A full pre-APR ensemble blocks 32 issues; when combined with post-APR review, the total rises to 37 blocked pull requests (72.5\%). However, as discussed in Section~\ref{sec:discussion}, this increases false positives and may be impractical in production settings.

Adding project test suites as a \emph{Fail}-based signal provides marginal benefit: +1 detection for noise attacks (both configurations) and +1 (Ensemble) / +2 (Best) for CVE reversion, with no gains for other attacks.

\subsection{Cost Analysis}
\label{sec:cost}

We quantify the economic asymmetry between attackers and defenders by comparing the cost of generating adversarial bug reports ($C_{\text{attack}}$), applying pre-APR counter-measures ($C_{\text{pre}}$), and executing APR ($C_{\text{apr}}$).

\noindent\textbf{Cost of the Attacker ($C_{\text{attack}}$).}
Generating adversarial bug reports is extremely inexpensive. Across all adversarial issues, the total cost of the generation pipeline was \$0.0133, yielding a mean cost of approximately \$0.000295 per bug report. Costs were estimated based on token usage and current pricing for a locally run Devstral model (\$0.1/M input tokens and \$0.3/M output tokens\footnote{Pricing based on OpenRouter \url{https://openrouter.ai/mistralai/devstral-small} and Mistral AI \url{https://mistral.ai/news/devstral}}). On average, this cost decomposes into input (\$0.000083 $\pm$ \$0.000040) and output (\$0.000212 $\pm$ \$0.000144) components.

\noindent\textbf{Cost of Pre-APR Defenses ($C_{\text{pre}}$).}
Pre-APR defenses incur higher costs while remaining only partially effective. Using unstructured outputs, the best-performing configuration (o4-mini) achieved a detection rate of 47.05\% at a mean cost of \$0.00268 $\pm$ \$0.00100 per checked issue; gpt-4.1-mini incurred a mean cost of \$0.00101 $\pm$ \$0.00041. Structured-output variants exhibited comparable costs (o4-mini: \$0.00284 $\pm$ \$0.00111; gpt-4.1-mini: \$0.00097 $\pm$ \$0.00027) without improving detection performance. Lightweight defenses such as PromptGuard, LlamaGuard, and Granite-Guardian were cheaper but detected fewer than 10\% of malicious issues.
% , limiting their practical utility.

\noindent\textbf{Cost of APR ($C_{\text{apr}}$).}
Executing APR dominates the cost profile. Across the evaluated projects, the mean APR cost per issue was \$0.87 $\pm$ \$0.39. Notably, in 90\% of cases, adversarial bug reports induced attacker-aligned patches, meaning this cost is frequently incurred to produce insecure or harmful changes.

\noindent\textbf{Summary.}
We observe an order-of-magnitude asymmetry. The cost of generating an adversarial bug report is on the order of $10^{-4}$ USD, while applying the most effective pre-APR defenses costs on the order of $10^{-3}$ USD per issue--approximately one order of magnitude higher---yet still fails to block most attacks. In contrast, APR execution costs on the order of $10^{0}$ USD per issue, roughly three orders of magnitude higher than $C_{\text{attack}}$ and two orders of magnitude higher than $C_{\text{pre}}$. This imbalance structurally favors attackers, who can generate adversarial inputs at negligible cost, while defenders incur substantially higher expenses both to filter malicious inputs and to execute APR:
\[
C_{\text{attack}} \;\approx\; 10^{-4}
\;\ll\;
C_{\text{pre}} \;\approx\; 10^{-3}
\;\ll\;
C_{\text{apr}} \;\approx\; 10^{0}
\quad \text{(USD/issue)}.
\]

\section{Discussion}
\label{sec:discussion}

In this section, we reflect on our empirical findings in relation to the research questions. We emphasize key patterns observed in the attacks, the effectiveness of deployed defenses, and broader implications for the security of APR pipelines.

\subsection{Adversarial Attacks for APR}
\label{attack_quality}

\begin{tcolorbox}[colback=lightgray, colframe=white, boxrule=0pt, sharp corners]
\textbf{RQ1:} Are APR systems prone to generating insecure or unintended code when given adversarial bug reports? \\
\textbf{Answer:} Adversarial bug reports can subvert APR pipelines at multiple stages, inducing the generation of insecure or malicious patches, triggering unintended behaviors during CI/CD execution, and wasting developer effort via plausible but irrelevant issues. In our study, 46 out of 51 crafted bug reports (90.2\%) led the APR system to produce patches consistent with the attacker’s objective, including CVE reversion, injection of vulnerable code, test-time execution, and semantically meaningless refactoring.
\end{tcolorbox}

\textbf{Quality of the issues from the human developer perspective.} All generated issues were analyzed by an experienced software developer and an experienced software security specialist. The reviewed issues appeared plausible and actionable for human developers, at least at first glance. After deeper investigation, most of them could be prevented by human. For example, Naive APR proposes fixes that are obviously not aligned with projects, or CI/CD and Inject attacks are obvious from a security perspective (``\textit{extract full environment to external server}'') and should immediately raise red flags although they present a plausible story. Other attack classes are more complicated. Noise APR, and revert CVE, are not that obvious from  the developer perspective. Our demonstration attacks do not exhaust all possibilities. To further elude detection, attacks may have malicious content split upon into multiple stages and thus make detection harder~\cite{przymus2025wolvesrepositorysoftwareengineering}. 

We attribute the high success rate of creating malicious issues to carefully structured bug reports. We guided the LLM to generate reports that mirror developer conventions, include code snippets, and provide project-specific context-factors known to improve APR success rates. This context acts as an anchor, helping the APR system localize and modify the target code more effectively.% \todo{this is actually in line with research of how AI can improve bug reports.  \url{https://conf.researchr.org/details/ease-2025/ease-2025-ai-models---data/14/Can-We-Enhance-Bug-Report-Quality-Using-LLMs-An-Empirical-Study-of-LLM-Based-Bug-Re}}

\textbf{Testcases deemed dangerous.}  Even APR systems that validate fixes via test execution may be exposed to security risks. Executing untrusted tests to assess patch plausibility can itself be an attack vector, especially if the test environment is not properly sandboxed. This reinforces the need for strict isolation and secure-by-design execution environments when deploying APR in practice.

\textbf{APR misalignment} arises even for vague or generic issues. For example, a minimally justified request to "revert recent changes" still prompts the system to synthesize a plausible revert patch, indicating reliance on surface cues rather than deep semantic validation.

We also evaluated attacks designed solely to generate developer burden through meaningless but compilable patches, e.g., arbitrary refactorings, redundant inlines, or unused imports. While such patches do not introduce functional errors, they impose unnecessary review and maintenance costs. %and may degrade long-term code quality.

Critically, the distinction between ``plausible'' and ``correct'' fixes remains unresolved in APR practice. Some adversarial reports included test cases, others did not---yet both yielded test-passing patches. This underscores a known limitation: passing unit tests does not guarantee semantic correctness, a fact long recognized in APR literature. 

Despite this, many systems claim real-world applicability by automatically generating pull requests from unvetted GitHub issues, overlooking malicious inputs. Our attacks are realistic, reproducible, and range from manual to fully automated generation. All bug reports use developer-consistent language and are submitted to private GitHub mirrors, demonstrating their plausibility in standard workflows. The breadth and effectiveness of these strategies expose critical weaknesses in current APR defenses, posing systemic risks to software integrity and supply chain security.

\subsection{Feasibility and Effectiveness of Defenses}

\begin{tcolorbox}[colback=lightgray, colframe=white, boxrule=0pt, sharp corners]
\textbf{RQ2:} What defenses are feasible, and at which points in
the workflow can attacks be mitigated effectively?\\ \textbf{Answer:} Effective mitigation of adversarial bug reports is feasible but non-trivial. Lightweight, structured LLM-based classifiers, e.g., o4-mini, offer practical pre-APR filtering with low false-positive rates. Runtime isolation in APR/CI environments can prevent exploitation, while post-APR analysis, e.g., Copilot, can flag issues, though often requiring manual review. Ensemble defenses improve detection rates but may incur unacceptable false positives. End-to-end protection demands combining targeted pre-filters, secure execution, and informed human oversight.
\end{tcolorbox}

\textbf{False-Positive Rates for Custom Prompts.}
Performing initial filtering using custom prompts is an efficient way of detecting potentially malicious requests. As Sommer et al.~\cite{sommer2010outside} mention, the feasibility of real-life usage of machine leaning techniques depends upon a low false-positive rate as otherwise analyst-fatigue will occur, leading to disabling of filtering mechanisms. To validate the false-positive rate, we selected a range of 100 existing issues of the \textit{psf/requests} project (issues \#6000--6100, excluding one deleted entry) and analyzed them analogous to our malicious issues.

Using structured output resulted in overall low false-positive rates. GPT-4.1-mini resulted in false-positives, while 4o-mini only resulted in a single one. Both of them tagged issue \#6063 (a security bypass inquiry). GPT-4.1-mini additionally tagged an issue (\#6075, containing only an external link). We argue that both cases are not real-false positives but potentially real-life malicious issues.

When using unstructured output, using 4o-mini resulted in 3 false-positives (issues \#6019, \#6059, \#6100, all valid). Using gpt-4.1-mini resulted in 34 false-positives, making the manual review-effort prohibitively high for real-world usage.

\textbf{Asymmetry of Costs.} Our attack (cost: $10^{-4}$ USD) leverages a fundamental asymmetry: it is cheap to execute  yet costly to defend. Adversaries can automatically generate realistic, context-aware bug reports with minimal effort. Using lightweight context retrieval and small LLMs, we construct prompts that produce plausible reports in a single pass---no expensive infrastructure required.

In contrast, defenses incur significant cost at multiple stages:
\textit{(1) Filtering (cost: $10^{-3}$ USD}) requires large LLMs or human triage. Since adversarial reports mimic legitimate submissions, simple spam filters or heuristics are ineffective. Low-volume, high-quality submissions evade rate limits and raise few red flags.
\textit{(2) APR execution (cost: $10^0$ USD)} is far more expensive than bug report generation. Models like SWE-agent rely on costly multi-turn interactions with large LLMs, often backed by human oversight or CI integration.
\textit{(3) Post-APR validation}, e.g., CodeQL, Copilot, manual review, adds further overhead, with costs accumulating even for patches that are ultimately rejected. This cost imbalance favors attackers: generating high-impact, low-effort reports is trivial, while defenders must invest heavily at each pipeline stage to detect and contain them.

\textbf{Defending against Adversarial Reports}
Pre-filter is helpful but insufficient. \textit{Revert} and \textit{Injection} attacks can be expressed entirely in natural language, and LLM-based APR systems infer patches even without explicit code. This becomes a cat-and-mouse problem, as attackers can easily rephrase to bypass filters. Distinguishing a malicious LLM-generated report from a legitimate one, e.g., produced by automated bug-finding tools, is extremely difficult. 
Post-APR defenses (test suites and post-filtering) improve detection, yet about 30\% of issues still slip through, while requiring significantly more resources than the attacker. 
Finally, human triage is already strained: many OSS maintainers report being overwhelmed by large volumes of LLM-generated bug reports even without adversarial intent~\cite{linaaker2024sustaining,slop_security_reports}. For these reasons, pre-filtering is a useful mitigation but cannot fully protect APR pipelines.

\subsection{Recommendations}

We suggest the following recommendations for new APR pipelines that integrate Large Language Models:

\begin{enumerate}[leftmargin=*, label=(\arabic*)]
    \item \textbf{Use Structured LLM Prompts.} Project-specific, structured-output prompts (e.g., o4-mini) provide cost-effective, low-noise filters for incoming issues.

\item \textbf{Integrate Runtime Isolation.} APR and CI/CD systems must assume adversarial inputs. Local or containerized execution, capped resources, and sandboxing prevent worst-case outcomes.

\item \textbf{Improve UX of Post-Detection Tools.} Copilot's hidden comments or vague warnings should be surfaced more clearly to support effective triage.

\item \textbf{Avoid Overreliance on Ensembles.} While pre- and post-APR ensembles blocked 63-68\% of attacks, they also raise false-positive risks, potentially harming developer trust.
\end{enumerate}

We stress the importance of \textbf{keeping humans in the loop} (HITL) to uphold safety and security. While copilot improves the efficiency of code and patch reviews, humans must review the resulting comments (including low-confidence comments) for their potential security impact. As illustrated in Section~\ref{attack_quality}, LLM-generated malicious patches are often easy to detect by humans thus preventing automated processing in the first place. The saved costs must be balanced with the time spent by humans on issue review.

\subsection{Guidance for Future Research}

By highlighting a novel class of threats in the software supply chain, our work urges the community to reconsider long-held trust assumptions in automated repair pipelines.

We advocate for a security-aware design of APR systems that balances repair effectiveness with resilience to adversarial misuse. While current systems prioritize passing test suites, future work should explore APR alignment strategies that enforce project-specific security policies during synthesis preventing behavior such as reverting known CVE fixes or disabling authentication logic.

Additionally, the community requires standardized adversarial benchmarks to complement functional suites like SWE-bench; these new benchmarks would formalize red-teaming evaluations to measure an APR agent’s resilience against prompt injection and context manipulation.

Finally, given that pre-filtering remains a ``cat-and-mouse'' challenge, researchers should investigate improving human-in-the-loop (HITL) workflows. This includes developing forensic tools that visualize patch provenance and highlight security-critical changes, thereby reducing reviewer fatigue and enabling more effective oversight of automated contributions.

\subsection{Ethical Implications}

Our work demonstrates that LLM-based APR systems can be manipulated into introducing malicious changes through adversarial bug reports. While this has clear security implications, our intent is to raise awareness of these risks and to enable the development of appropriate defenses. The prototype developed for this study and as well as all artifacts are be open-source, alongside a curated replication package to support further research and transparency.

We acknowledge the abuse potential of automated vulnerability injection and, following responsible-disclosure practices and prior guidance on LLM misuse~\cite{greshake2023youvesignedforcompromising}, notified affected stakeholders, particularly GitHub. All experiments were conducted on forked repositories under controlled conditions, with no interaction with production systems.

\section{Threats to Validity}
\label{sec:threats-to-validity}

\textbf{Internal Validity.} Our manual annotation process involves expert judgment, which may introduce subjectivity. To mitigate this, we used a dual-annotator protocol with independent assessments and discussion-based resolution. Still, subtle vulnerabilities or security implications may have been overlooked. LLMs are inherently non-deterministic. Typically this threat is mitigated by fixing the sampling temperature to 0. We were not able to perform this, as deviating from the default settings of the used APR system and utilized defenses negatively impacts Construct Validity.

\textbf{External Validity.}
Our experiments focus on a subset of open-source projects and a single APR system (\textit{SWE-agent}). While the selected repositories are diverse and representative of real-world codebases, our findings may not generalize to proprietary software or APR tools with substantially different architectures or training regimes. However, our adversarial methodology exploits structural properties shared by many modern APR systems: reliance on unvetted natural language input, automated patch generation via LLMs, and validation based primarily on test-suite outcomes. These properties are common to other state-of-the-art systems~\cite{yang2024sweagent,bouzeniaRepairAgentAutonomousLLMBased2024,zhang2024autocoderover, wang2025openhandsopenplatformai}. Unless such tools are explicitly hardened with security-aware defenses--a feature currently absent in most publicly available APR pipelines -- we expect similar vulnerabilities to manifest.

\textbf{Construct Validity.}
We define attack success in terms of patch generation, detection bypass, and the presence of malicious behavior. These metrics are practical, but may not fully capture downstream impact, e.g., exploitation feasibility or user-facing harm. %Similarly, false positives in filter models or review tools could influence results.

\textbf{Tool Limitations.}
Our work relies on specific versions of LLMs (e.g., GPT-4o), static analyzers (CodeQL), and review tools (Copilot). Changes in these tools or deployment environments may affect replicability. Additionally, APR behavior can be sensitive to prompt phrasing and model configuration.

\section{Related Work}

\subsection{Automated Program Repair (APR)}

Recent work has shown that large language models can drive automated program repair through autonomous agents. Systems like RepairAgent and AutoCodeRover demonstrate that LLM-based agents can triage issues, modify code, and validate changes in real-world repositories \cite{bouzeniaRepairAgentAutonomousLLMBased2024, zhang2024autocoderover}. Similarly, SWE-agent introduces an Agent-Computer Interface (ACI) enabling agents to navigate and act within codebases, achieving state-of-the-art results on the SWE-bench benchmark~\cite{yang2024sweagent}. However, these systems prioritize performance and reliability, with little consideration for security. They do not address whether adversaries could manipulate agents to introduce insecure or malicious changes.

Benchmark suites like SWE-bench and SWE-bench Multimodal assess the functional correctness of LLM-generated patches~\cite{jimenezSWEbenchCanLanguage2024a, yangSWEbenchMultimodalAI2024}, but overlook whether those patches preserve security properties or introduce vulnerabilities.
While security risks of LLMs are well documented (e.g., prompt injection, unsafe code~\cite{lyuAutomaticProgrammingLarge2024}), they remain largely unexplored in LLM-based APR.

Our work fills this gap by investigating the security risks of LLM APR systems, demonstrating that adversarial bug reports can reliably induce unsafe behavior, even in state-of-the-art LLM agents.

\subsection{LLM Safety and Security}

Agentic AI systems, such as APR agents, are high-value targets for attackers, as they are positioned at the intersection of private data, untrusted user content, powerful actions that an be invoked, and external communication~\cite{trifecta}.

Bugdar~\cite{naulty2025bugdaraiaugmentedsecurecode} is an AI-augmented secure code review tool that integrates directly into GitHub pull requests. It combines fine-tuned LLMs with Retrieval-Augmented Generation to provide contextual, project-specific vulnerability analysis across multiple languages (e.g., Solidity, Rust, Python).

\subsubsection{Attacks Against LLM Applications}
%\label{background:attacks}

The \textit{OWASP Top 10 for LLM Applications}~\cite{owasp_top10} highlights common attacks and issues occurring in deployed LLM applications. We ground our research on this established industry standard to make our results applicable and relevant for real-world APR systems.

\textit{Prompt Injection} attacks (LLM01) were the top threat and will also be used within our evaluation. The impact of these attacks is often \textit{Sensitive Information Disclosure} (LLM06) or \textit{Insecure Output Handling} (LLM02). The latter includes LLMs executing unintended commands and is often amplified through \textit{Excessive Agency} (LLM08) by LLMs. If results of a LLM are used or incorporated without human oversight, \textit{Overreliance} (LLM09) occurs.

We decided to not employ \textit{Training Data Poisoning} (LLM03) or utilize \textit{Supply Chain Vulnerabilities} (LLM05), do not exploit \textit{Insecure Plugin Design} (LLM07), 
or
%\textit{Model Denial of Service} (LLM04) nor
\textit{Model Theft} (LLM10).

\subsubsection{Defenses for LLM-based Applications}

%Based on our analysis of potential attacks (Section~\ref{background:attacks}), 
We group defenses info measures that try to prevent Prompt Injection attacks, and measures that try to minimize the impact of a successful attack.

Defenses against Prompt Injection Attacks are commonly based on filtering user-provided prompts before they are interpreted by the LLM utilized by APR. This filtering is typically performed by specialized LLMs, fine-tuned for detecting malicious content, that are cheaper to operate due to their reduced model parameter count. Examples of this are Meta LlamaGuard~\cite{inan2023llama} for detecting malicious intend within prompts, or Meta PromptGuard~\cite{prompt_guard} for detecting Jailbreak and Prompt Injection attacks. Google CaMeL~\cite{debenedetti2025defeating} employs techniques inspired by \textit{Control-Flow-Integrity} to enforce separation between user-provided data and program data (prompts). AlignmentCheck~\cite{chennabasappa2025llamafirewall} monitors multi-step LLM trajectories to verify that the task currently performed by a model and the user's original goal are staying in alignment.

The result of APR agents is typically a patch. To reduce the impact of a malicious patch, traditional counter-measures such as Static Application Security Testing (SAST) can be employed. Meta CodeShield~\cite{chennabasappa2025llamafirewall} is a static-analysis engine designed to validate LLM-generated code and explicitly states protection against \textit{malicious code via prompt injection} as its goal.

\section{Conclusion}

As APR systems increasingly rely on natural language and large language models, they inherit a new class of threats stemming from unvetted user input. This paper demonstrates that adversarial bug reports---crafted to appear plausible while embedding malicious intent---pose a significant and underrecognized risk to the security and integrity of APR.

Our empirical study shows that these adversarial inputs can systematically subvert even state-of-the-art APR pipelines: in our demonstration, over 90\% of the crafted bug reports resulted in patches aligned with the attacker's objective, including CVE reversions, vulnerability injection, and CI/CD abuse. Pre- and post-APR defenses, while helpful, often failed silently (best filter only blocked 47\% issues) or demanded high levels of manual effort and scrutiny (Co Pilot supported review 58\%). Project test suites only slightly improve attack detection.
The fundamental asymmetry---low-cost attack versus high-cost defense---further compounds the challenge.

We contribute a modular framework for generating and testing adversarial bug reports, enabling scalable and reproducible evaluations of APR security. Our findings reveal that naive trust in bug reports, even in sophisticated repair systems, creates a structural vulnerability that is ripe for exploitation.

To mitigate this threat, we recommend a multi-pronged approach: lightweight, structured LLM-based input filtering; secure-by-default CI environments; more transparent post-repair validation tooling; and provenance tracking of APR-generated patches. Red-teaming and adversarial robustness testing should become standard practice for any deployment of automated repair agents.

Ultimately, this work calls for a shift in how the community thinks about trust, automation, and security in APR workflows. As LLMs gain autonomy in the software supply chain, we must move beyond accuracy benchmarks and confront their adversarial brittleness head-on.

\balance
\bibliographystyle{ACM-Reference-Format}
\bibliography{bibliography}

\end{document}